\documentclass[twocolumn,amsmath,amssymb,english,aps,prb,floatfix,showpacs]{revtex4}
\usepackage{times}
\usepackage[english]{babel}
\usepackage{graphicx}



\def\ket#1{\left|#1\right>}

\begin{document}

\title{Kondo effect in double quantum dots with inter-dot repulsion }

\author{J. Mravlje$^{1}$, A. Ram\v{s}ak$^{2,1}$, and T. Rejec$^{1,2,3}$}

\affiliation{$^{1}$Jo\v{z}ef Stefan Institute, Ljubljana, Slovenia}

\affiliation{$^{2}$Faculty of Mathematics and Physics, University of Ljubljana,
Slovenia}

\affiliation{$^{3}$Department of Physics, Ben-Gurion University, Beer-Sheva,
Israel}

\begin{abstract}
We investigate a symmetrical double quantum dot system serially attached
to the leads. The emphasis is on the numerical analysis of finite
inter-dot tunneling in the presence of inter-dot repulsive capacitive
coupling. The results reveal the competition between extended Kondo
phases and local singlet phases in spin and charge degrees of freedom.
 The corresponding phase diagram is determined quantitatively. 
\end{abstract}

\pacs{73.23.-b, 73.63.Kv, 72.15.Qm}

\maketitle
Quantum dots\cite{reimann02,kemerink94,blick98,wiel03} provide in
addition to various applications in proposed spintronics\cite{zutic04}
and quantum computation\cite{nielsen01,diVincenzo05} devices a
playground for studying phenomena known in bulk condensed-matter
systems. Specifically, the Kondo effect was found to play an important
role in single\cite{goldhaber-gordon98} and double quantum dot
\cite{heong01,wilhelm02,holleitner02} (DQD) systems. From early
theoretical work on the low-temperature properties of the two-impurity
Kondo Hamiltonian \cite{jones88,jones89} it is known that localized
moments form either each its own Kondo singlet with delocalized
electrons in the leads -- \emph{double Kondo} phase (2K) -- or they
form \emph{local spin-singlet} state (LSS) decoupled from delocalized
electrons. The crossover between the two regimes is the consequence
of competing energies of Kondo singlet with Kondo temperature $T_{K}$ 
and of  LSS formation $J$. Similar results were
obtained by the analysis of a two-impurity Anderson model by means of
slave-boson formalism \cite{aono98,georges99,aguado00,aono01,lopez02} and
numerical renormalization group (NRG)\cite{izumida99,izumida00}.
Resembling behavior was found also in particular regimes of a triple
quantum dot system\cite{zitko06}.

Here we focus on the role the inter-dot (capacitive) interaction plays
in the electron transport through serially coupled DQD. The influence
of the capacitive coupling was already studied by means of the
equation-of-motion method \cite{you99,lamba00,busser00,bulka04} which,
however, fails to capture the Kondo correlations accurately. On the
other hand, the Kondo correlations were considered in the limit of
vanishing tunnel coupling and strong inter-dot interaction where the
ground state of the isolated DQD exhibits four-fold degeneracy: in
addition to spin degeneracy also singly occupied states labeled with
$(1,0)$ and $(0,$1) are degenerate leading to orbital Kondo behavior
\cite{pohjola01,borda03,sakano04,sakano05,chudnovskiy05}. Here
$(n_{1},n_{2})$ corresponds to occupancies $n_{1}$ and $n_{2}$ of the
two dots.  The simultaneous presence of spin and orbital degeneracy
results in enhanced Kondo temperature
$T_{K}(N)\sim\sqrt{J'\rho_{0}N}\exp(-1/J'\rho_{0}N)$, where $N$ is the
degeneracy, $\rho_{0}$ the noninteracting density of states, and $J'$
the corresponding Schrieffer-Wolf prefactor \cite{borda03}.
\begin{figure}
\includegraphics[%
  width=75mm]{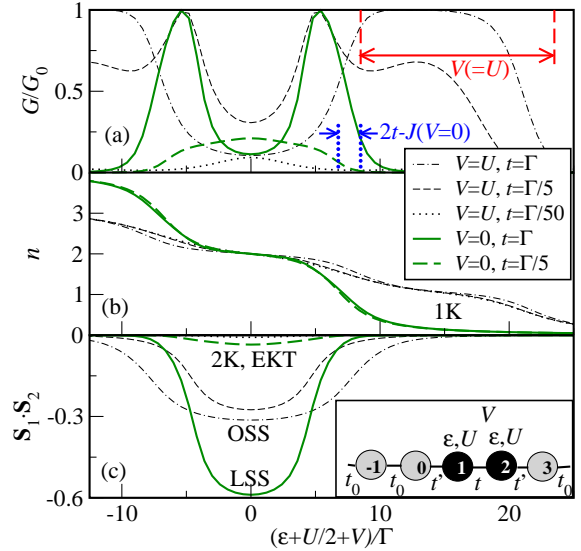}
\caption{(Color online) \label{cap:Fig1}Conductance (a), occupancy (b) and
spin-spin correlation (c) of DQD for $V=0$ (thick), and $V=U$ (thin). The vertical lines for $V=U$
(dashed), and $V=0$ (dotted) correspond to transitions of isolated DQD
between $n=0,
n=1$ and $n=1, n=2$ for $t=\Gamma$. Spin-singlet regimes (LSS, OSS)
and  Kondo regimes (2K, EKT, 1K) are indicated with labels.
Inset: DQD system with both: inter- and intra-dot Coulomb coupling,
attached to noninteracting leads.}
\end{figure}
The isospin Kondo resonance was observed indirectly in the
measurements of enhanced conductance through
DQDs\cite{heong01,wilhelm02,holleitner02,sasaki04} and carbon
nanotubes\cite{herrero05, aguado05} and also directly, in the bulk,
with techniques of the scanning tunneling
microscopy\cite{kolescnicko02}.  Recently the \emph{enhanced Kondo
temperature} phase (EKT) was predicted for DQD with inter-dot
interaction $V=U$ in the absence of tunneling between the dots also at
half-filling\cite{krishnamurthy05}, i.e., $n=n_{1}+n_{2}=2$. The main
characteristics of this phase at low temperatures is the enhanced
width of the Kondo resonance (i.e., Kondo temperature $T_K$) on top of
an incoherent continuum in the density of states. However, the
analysis of capacitively coupled DQD with finite inter-dot tunneling
rates has been lacking to date and we address this issue in the
present paper.

We model DQD shown in inset of Fig.~\ref{cap:Fig1}(c) by the Anderson-type
Hamiltonian $H=H_{\mathrm{d}}+H_{\mathrm{l}}$, where $H_{\mathrm{d}}$
corresponds to the isolated dots \[
H_{\mathrm{d}}=\sum_{i=1,2}(\epsilon n_{i}+Un_{i\uparrow}n_{i\downarrow})+Vn_{1}n_{2}-t\sum_{\sigma}(c_{1\sigma}^{\dagger}c_{2\sigma}+h.c.),\]
with $n_{i}=n_{i\uparrow}+n_{i\downarrow}$, $n_{i\sigma}=c_{i\sigma}^{\dagger}c_{i\sigma}$.
The dots are coupled by a tunneling matrix element $t$ and a capacitive
$V$ term. The on-site energies $\epsilon$ and the Hubbard repulsion
$U$ are taken equal for both dots. $H_{\mathrm{l}}$ corresponds
to the noninteracting left and right tight-binding lead and to the
coupling of DQD to the leads, \[
H_{\mathrm{l}}=-t_{0}\sum_{i\neq 0,1,2}c_{i\sigma}^{\dagger}c_{i+1\sigma}-t'\sum_{\sigma}(c_{0\sigma}^{\dagger}c_{1\sigma}+c_{3\sigma}^{\dagger}c_{2\sigma})+h.c.,\]
 where the sites are labeled as shown in the inset of Fig.~\ref{cap:Fig1}(c). 

In this paper we study the low temperature properties of DQD, derived
from the ground state at $T=0$, determined by the Gunnarsson and
Sch\"{ o}nhammer projection-operator
method\cite{gunnarsson85,rr03}. The conductance is calculated using
the sine formula \cite{rr03},
$G=G_{0}\sin^{2}[(E_{+}-E_{-})/4t_{0}L]$, where $G_{0}=2e^{2}/h$ and
$E_{\pm}$ are the ground state energies of a large auxiliary ring
consisting of $L$ non-interacting sites and an embedded DQD, with
periodic and anti-periodic boundary conditions, respectively. The
chemical potential is set in the middle of the band, which corresponds
to $L$ electrons in the ring. The method proved to be very efficient for
various systems with Kondo correlations
\cite{rr03,rbr04,jm05,zitko06}. Up to $N\sim5000$ sites were used with
variational ansatz based on $\sim 100$ projection operators, which was
sufficient to obtain converged results for the boundaries between
various regimes.

For large inter-dot tunneling $t\gtrsim U$ molecular bonding and
anti-bonding orbitals with energies $\epsilon_{b(a)}=\epsilon\mp t$
are formed. Whenever either of the two orbitals is singly occupied, the physics is analogous to
the single impurity Anderson problem with unitary transmission and 
Kondo correlations (1K regime). When the bonding orbital
is doubly occupied the electrons form an \emph{orbital spin-singlet}
state (OSS) with diminished conductance as the Kondo peak in the
density of states is absent. 
The other limit, $t\to0$,
is more interesting. When also inter-dot repulsion is moderate $V\lesssim U$
the dots are effectively decoupled and in the particle-hole symmetric
point ($\epsilon+U/2+V=0$) the 2K state occurs.
Regimes with different occupancies are separated by the intermediate
valence regime of width determined by the noninteracting hybridization
$\Gamma=t'^{2}/t_{0}$. 

Behavior of the system is less obvious in the  intermediate region
$0\lesssim t\lesssim U$. In Fig.~\ref{cap:Fig1}(a-c) we plot the conductance, occupancy and
spin-spin correlation for $V=0$
and $V=U$ ($\Gamma=t_{0}/25$ and $U/\Gamma=15$ is kept constant
throughout the paper). The curves for $t=\Gamma, V=U$ (dashed-dotted
thin lines) are
the most reminiscent of known results for single impurity case. The
wide plateau in conductance curve 
corresponds to single occupancy regime. While the
occupancy for the case with reduced $t=\Gamma/5$ (dashed thin line) does not change significantly,
two maxima in conductance become
discernible. One reflects the 1K Kondo physics corresponding to the
bonding orbital ($n=1$ there). The other maximum (with unitary
conductance) is known from the 
$V=0$ case\cite{georges99,izumida00} and occurs generally at the
transition between extended Kondo and decoupled spin-singlet states
at $n\sim2$. For smaller $t=\Gamma/50$ at $n\sim 1$ conductance is
significantly reduced, but the ground state is still 1K, which is signaled with the
remaining plateau in occupancy and diminished fluctuations of
the occupation of the relevant, (\emph{i.e.}, bonding or anti-bonding)
orbital (not shown here). In this case the OSS state is not
energetically favorable near $n\sim2$, correspondingly the conductance peak
with the unitary conductance is absent (likewise for $V=0$,
$t=\Gamma/5$, where the LSS is not energetically preferred). At $n=2$
rather the EKT phase is found, as discussed later. 

The $V=0$ case is seen to differ qualitatively from the $V=U$ case. Crude
insight into distinction between the two may be gained from treating just the
filling properties of a detached system \cite{harris67}. The first
electron is added when $\epsilon=t$, and the second when
$\epsilon=-t+J+[(U+V)-|U-V|]/2$, where $J=[-|U-V|+\sqrt{(U-V)^{2}+16t^{2}}]/2$ is the difference
between singlet and triplet energy of isolated DQD. When $n=2$ the
ground state is $[\alpha
(\ket{\uparrow\downarrow}-\ket{\downarrow\uparrow}) +
\beta(\ket{20}-\ket{02})]/\sqrt{2}$, where $\alpha/\beta =
4t/(V-U+\sqrt{(U-V)^2+16t^2})$. 
For $V=0$ and $V=U$ case the values of $\epsilon$ where the occupancy
of a ground state changes from $n=0$ to $n=1$ and from $n=1$ to $n=2$
are indicated by vertical lines on Fig.~\ref{cap:Fig1}(a). The
range of $\epsilon$ where single occupation is favorable is
diminished in the $V=0$ case. 
\begin{figure}[t]
\includegraphics[%
  width=75mm]{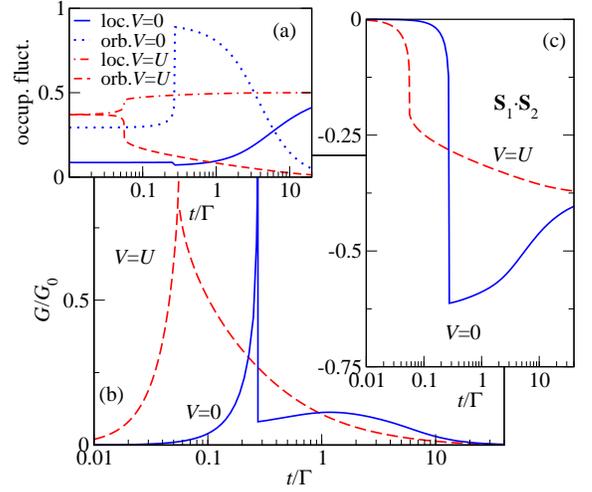}
\caption{\label{cap:Fig2}(Color online) (a) Orbital $\Delta n_{b}^{2}=\Delta
  n_{a}^{2}$ and site $\Delta
  n_{1}^{2}=\Delta n_{2}^{2}$ occupancy fluctuations   at half-filling $(n=2)$ as a function
of $t/\Gamma$ for $V=0$ and $V=U$. Conductance (b) and spin-spin
  correlation (c), for $V=0$ (full) and $V=U$ (dashed). 
}
\end{figure}

This observation remains valid also for DQD
attached to the leads, Fig.~\ref{cap:Fig1}(b). At first sight surprising feature is seen in
Fig.~\ref{cap:Fig1}(c), where the expectation value of a
spin-spin correlator $\mathbf{S}_1\cdot\mathbf{S}_2$ is seen
to approach $-3/8$ near $n=2$, but in fact it is just equal to the noninteracting
result with two electrons occupying the bonding orbital. The fact that the
spin-spin correlator approaches $-3/4$ near $n=2$ in the $V=0, t=\Gamma$ case suggests that
the local picture prevails. The electrons indeed form
singlet in the local basis what is seen also in diminished value of
the local occupancy fluctuations, Fig.~\ref{cap:Fig2}(a).
 \textsf{}Additional insight into precise role of inter-dot interaction can be obtained
by re-writing the total Hamiltonian in the basis of orbital operators,

\begin{eqnarray}
H_{\mathrm{d}} & = & \sum_{\alpha=a,b}\left[\epsilon_{\alpha}n_{\alpha}+\frac{U+V}{2}\left(n_{\alpha\uparrow}n_{\alpha\downarrow}+n_{\alpha\uparrow}n_{\bar{\alpha}\downarrow}\right)\right]+\nonumber \\
 & & +  V\sum_{\sigma}n_{a\sigma}n_{b\sigma}+\frac{U-V}{2}
\left(C_{\textrm{flip}}-S_{\textrm{flip}}\right)\nonumber,\end{eqnarray}
where notation $\bar{a}=b$, $\bar{b}=a$ is used. The last term of
$H_{\mathrm{d}}$ consists of charge-flip $C_{\textrm{flip}}=T_{a}^{+}T_{b}^{-}+h.c.$
and spin-flip $S_{\textrm{flip}}=S_{a}^{+}S_{b}^{-}+h.c.$ operators,
where $S_{\lambda}^{-}=c_{\lambda\downarrow}^{\dagger}c_{\lambda\uparrow}=(S_{\lambda}^{+})^{\dagger}$
are spin and $T_{\lambda}^{-}=c_{\lambda\uparrow}c_{\lambda\downarrow}=(T_{\lambda}^{+})^{\dagger}$
charge (isospin)\cite{taraphder91} lowering and raising operators
for the orbitals $\lambda=b,a$ (or sites $\lambda=1,2$). The full
spin (isospin) algebra is closed with operators $S_{\alpha}^{z}=(n_{\alpha\uparrow}-n_{\alpha\downarrow})/2$
and $T_{\alpha}^{z}=(n_{\alpha}-1)/2$, respectively. 

When $V=U$, the spin- and isospin-flip terms in $H_{\mathrm{d}}$
are absent: the Hamiltonian is mapped exactly to the two-level Hamiltonian
with intra- and inter-level interaction $U$ with the bonding and
anti-bonding levels coupled to even and odd transmission channels,
respectively. When $V\neq U$ this mapping is no longer strictly
valid. Taking $V=0$ case as an extreme example, we find two mechanisms
responsible for this. Firstly, 
the electrons can avoid the inter-level repulsion by
occupying aligned spin-states in different orbitals, and secondly,
charge flip-terms induce the fluctuations of  charge between
orbitals. Both mechanisms prohibit electrons to occupy the well-defined orbital
states. 

In the rest of the paper we concentrate on the symmetric
point of the model, $n=2$. The manifestation
of adequacy of  orbital picture are the fluctuations
of occupancy $\Delta
n_\lambda^2=(n_{\lambda}-\left<n_{\lambda}\right>)^2$, shown in 
Fig.~\ref{cap:Fig2}(a). For $V=U$ orbital occupancy fluctuations  $\Delta
n_b^{2}$(= $\Delta n_a^{2}$ for $n=2$) are indeed smaller than local occupancy
fluctuations  $\Delta
n_1^{2}=\Delta n_2^{2}$ for all
values of $t$. On contrary, for $V=0$, $\Delta
n_b^{2}<\Delta n_1^2$ only for $t\gtrsim 5U$. 


\begin{figure}[b]
\includegraphics[%
  width=68mm]{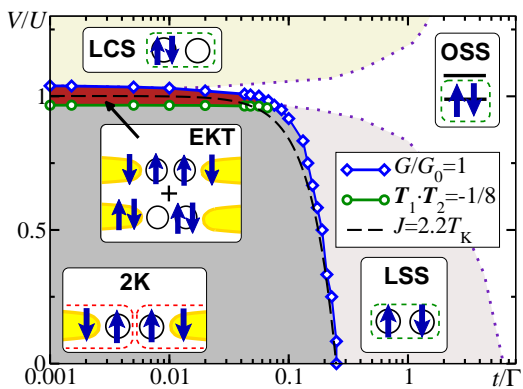}
\caption{(Color online) \label{cap:Fig3}Phase diagram in ($t/\Gamma,V/U)$
plane for $n=2$. }
\end{figure}

To explain the conductance and spin-spin correlation shown in Fig.~\ref{cap:Fig2}(b,c) we
first note that the electrons bind into singlet (local or orbital, depending
on the $V/U$), whenever $J\gtrsim J_c=2.2T_{K}$, 
which is known for $V=0$  results\cite{su2,jones88}, but 
we find that this result can
be readily generalized for by taking the difference between
singlet and triplet energies $J$ also for $V>0$.
This relation indicates that although the Kondo temperature is known
to rise in the $V=U$ case\cite{krishnamurthy05} for $n=2$, the
OSS is energetically favorable for moderate $t$ as $J(V\sim U)\propto t$
is larger when compared to $J(V=0)\sim 4t^{2}/U$. 

It should be noted that due to the variational nature of the method,
the boundaries between various regimes are accurately
reproduced. Sharp transitions of correlation functions in the vicinity
of crossovers are less precisely determined. The transitions
discussed here are actually smoothened into crossovers what happens
generally when the parity breaking terms (here the inter-dot
tunneling) are present in the Hamiltonian \cite{sakai92}.

For $V=0$ in absence of inter-dot tunneling $t\to0$, each
of the dots forms a Kondo singlet with its own lead (2K regime). When
$t$ is increased above some value $t_c$, set by $J\gtrsim J_c$,
the ground-state becomes a LSS, what is seen from increased value
of $-\mathbf{S}_{1}\cdot\mathbf{S}_{2}\to3/4$. The conductance is
small in both cases due to the effectively decoupled leads in the
2K case or deficiency of states near the chemical potential in the
LSS case. The conductance $G\to G_{0}$ at some point what is most
vividly understood in terms of the phase shift between odd and even transmission
channels, changing from 0 to $\pi$ during the transition\cite{georges99}.
For $V\lesssim U$ the behavior does not change qualitatively, as
the transition remains close to $2T_{K}\sim J$. In the case $V\sim U$
numerical results show deviations from this prediction: the transition
occurs at smaller values of $t$ than set with relation $J\gtrsim J_c$.
We explain this by increased Kondo temperature in the
EKT phase as suggested from NRG
results \cite{krishnamurthy05}. Here the transition between 'SU(4)
Kondo' \cite{su4} and OSS occurs at $T_{K}^{SU(4)}\sim J(V=U)=2t_c^{*}$.

Further increasing $V$ above $U$ we find, that in contrast to the strict $t=0$
case where the ground state for $V\gtrsim U$ is a non-Fermi liquid
charge-degenerate phase\cite{krishnamurthy05}, a finite $t$ breaks the
degeneracy resulting in a (Fermi liquid) phase with increased magnitude of isospin-isospin
correlation $\mathbf{T}_1\cdot\mathbf{T}_2$: a \emph{local charge
  singlet} (LCS). LCS corresponds to the ground state of a detached
system for $(V-U)/t\gg1$.

\begin{figure}[t]
\includegraphics[%
  width=75mm]{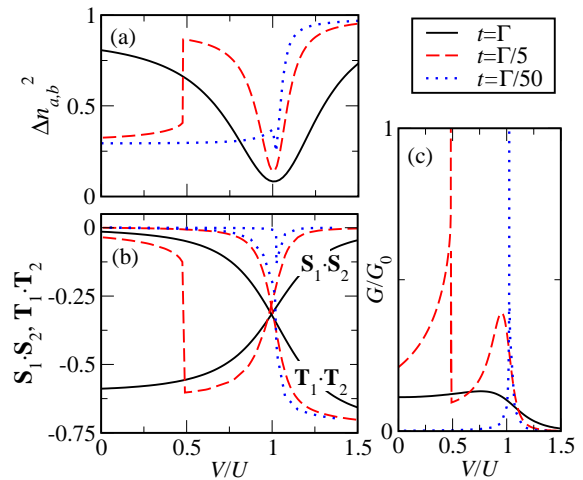} 
\caption{(Color online) \label{cap:Fig4} (a) Fluctuations of charge in orbitals
$\Delta n_{a}^{2}=\Delta n_{b}^{2}$, (b) the spin-spin $\mathbf{S}_{1}\cdot\mathbf{S}_{2}$
and isospin-isospin correlation $\mathbf{T}_{1}\cdot\mathbf{T}_{2}$
are plotted for $t>t_c$ (full), for $t_c^{*}<t<t_c$
(dashed) and $t<t_c^{*}$ (dotted). (c) The conductance
for the three cases. }
\end{figure}

In Fig.~\ref{cap:Fig3} the phase diagram of DQD for $n=2$ is shown.
The border line between Kondo and local-singlet phases (full line
with diamonds) is characterized by unitary conductance and the abrupt
increase of $-\mathbf{\mathbf{S}}_{1}\cdot\mathbf{S}_{2}$ for the
transition between 2K and LSS and of $-\mathrm{\mathbf{T}}_{1}\cdot\mathbf{T}_{2}$
for transition between EKT and LCS {[}see also Fig. \ref{cap:Fig2}
(a,b) and Fig. \ref{cap:Fig4}(b,c){]}.  
In the $t,t'\to0$ limit the states
$(2,0),(0,2)$ and four states $(1,1)$ are degenerate, and $\left<-\mathrm{\mathbf{T}}_{1}\cdot\mathbf{T}_{2}\right>=1/6$,
if the expectation value is taken with respect to the ground state
where each of these states occurs with equal probability. We quantitatively
characterize
the transition to the EKT phase (full line with circles) when $-\mathrm{\mathbf{T}_{1}}\cdot\mathbf{T}_{2}>1/8$.
The border lines which we use to distinguish between local and orbital
regimes (dotted) are given by $\Delta n_{b}^{2}=\Delta n_{1}^{2}$ 
for $V<U$ and by $-\mathbf{\mathbf{S}}_{1}\cdot\mathbf{S}_{2}=3/16$
(half of noninteracting value) for $V>U$. %

In Fig.~\ref{cap:Fig4} we sample the phase diagram at different
$t$ by increasing $V$. We plot three different cases. In the first case
(full curve) DQD is in LSS already at $V=0$. By increasing $V$ we
see a crossover to LCS through the orbital regime. In the other two
cases we start in 2K. Now if $t<t_{c}^{*}$ (dotted) we stay in the
Kondo regime until the transition to LCS happens with the unitary
conductance at the transition point. However if $t>t_{c}$ (dashed)
there is an area in between where LSS is energetically favorable.
From the point where the system is in LSS, it follows the same scenario
as in the first case.

Finally, let us briefly comment on $V>U$ part of the phase-diagram. This
regime could arise when the dots are capacitively coupled while the
on-site repulsion is reduced due to the coupling to local Einstein
vibrational modes \cite{hewson80,jm05}. Here the competition between
LCS and EKT is relevant for the ground-state. Our numerical data confirms
NRG predictions that the Kondo regime is relevant only for small increase
of $V$ over $U.$ For parameters used here, we find critical $V_{\mathrm{c}}\sim U+U/20.$ 

In conclusion, we have presented the phase diagram of a pair of tunneling-coupled
quantum dots with Coulomb inter- and intra-dot interaction. In general,
the behavior of $V\sim U$ case was found to differ from the $V\sim0$
case due to different arrangement of  orbital levels of a decoupled system.
 In addition, for $V\sim U$ the correct exchange coupling is not $J\sim 4t^{2}/U$ but rather
$J\sim2t$, hence the spin-singlet binding energy is increased. The
increase in $J$ is seen to surmount the enhancement of Kondo temperature
 which is due to the additional degeneracy in the charge sector. Therefore this
enhanced Kondo effect for $n=2$ should be discernible in
measurements only in a tiny window of tunneling rates and inter-dot
vs. intra-dot interaction.

We thank R. \v{Z}itko for valuable discussions and
acknowledge the support of MSZS under grant Pl-0044.


\begin{thebibliography}{10}
\bibitem{su4} SU(4) Kondo is seen also in increased value of 
$-\mathbf{T}_{1}\cdot\mathbf{T}_{2}$, but with simultaneously $\mathbf{S}_{1}\cdot\mathbf{S}_{2}\sim 0$.
\end{thebibliography}
\end{document}